\documentclass[lettersize,journal]{IEEEtran}
\usepackage{amsmath,amsfonts}
\usepackage{algorithmic}
\usepackage{algorithm}
\usepackage{bm}
\usepackage{color, xcolor}
\usepackage{array}
\usepackage[caption=false,font=normalsize,labelfont=sf,textfont=sf]{subfig}
\usepackage{textcomp}
\usepackage{stfloats}
\usepackage{url}
\usepackage{verbatim}
\usepackage{graphicx}
\usepackage{cite}
\usepackage{amssymb}  %后加的
\usepackage{multirow}
\usepackage{makecell}
\usepackage{framed} 
\renewcommand{\bf}{\bm}

\begin{document}

\title{Interleaved Block-Sparse Transform}

\author{Lei~Liu,~\IEEEmembership{Senior Member,~IEEE}, Ming~Wang, Shufeng~Li,~\IEEEmembership{Member,~IEEE}, Yuhao~Chi,~\IEEEmembership{Senior Member,~IEEE}, Ning~Wei, and ZhaoYang~Zhang,~\IEEEmembership{Senior Member,~IEEE}
% \thanks{This work was supported by ZTE Industry-University-Institute Cooperation Funds under Grant No.IA20231213009.}
\thanks{Lei Liu and Zhaoyang Zhang are with the Zhejiang Provincial Key Laboratory of Information Processing, Communication and Networking, College of Information Science and Electronic Engineering, Zhejiang University, Hangzhou, 310007, China (e-mail: \{lei\_liu, ning\_ming\}@zju.edu.cn).}
\thanks{Ming Wang and Shufeng Li are with the State Key Laboratory of Media Convergence and Communication, School of Information and Communication Engineering, Communication University of China, Beijing 100024, China (email: \{wangming12136, lishufeng\}@cuc.edu.cn).}
\thanks{Yuhao Chi is with the State Key Laboratory of Integrated Services Networks, School of Telecommunications Engineering, Xidian University, Xi’an, 710071, China (e-mail: yhchi@xidian.edu.cn).}
\thanks{Ning Wei is with the ZTE Corporation and also with the State Key Laboratory of Mobile Network and Mobile Multimedia Technology, Shenzhen, 518055, China (e-mail: wei.ning@zte.com.cn).}}
\maketitle

\begin{abstract}
Low-complexity Bayes-optimal memory approximate message passing (MAMP) is an efficient signal estimation algorithm in compressed sensing and multicarrier modulation. However, achieving replica Bayes optimality with MAMP necessitates a large-scale right-unitarily invariant transformation, which is prohibitive in practical systems due to its high computational complexity and hardware costs. To solve this difficulty, this letter proposes a low-complexity interleaved block-sparse (IBS) transform, which consists of interleaved multiple low-dimensional transform matrices, aimed at reducing the hardware implementation scale while mitigating performance loss. Furthermore, an IBS cross-domain memory approximate message passing (IBS-CD-MAMP) estimator is developed, comprising a memory linear estimator in the IBS transform domain and a non-linear estimator in the source domain. Numerical results show that the IBS-CD-MAMP offers a reduced implementation scale and lower complexity with excellent performance in IBS-based compressed sensing and interleave frequency division multiplexing systems.  
\end{abstract}


\section{Introduction}
In signal processing and wireless communication applications, a standard linear system is widely employed to recover an unknown signal $\bf{s}$ from a noisy observation $\bf{y}$, i.e., 
\begin{equation}\label{Eqn:sls}
    \bf{y} = \bf{A\Xi s} + \bf{n},
\end{equation}
where ${\bf{\Psi =A\Xi}}\in {\mathbb{C}}^{M\times N}$ is the measurement matrix and ${\bf{n}}$ is an additive white Gaussian noise (AWGN). In various applications, $\bf{A}$ and $\bf{\Xi}$ serve different functional roles. To effectively support these applications, it is critical to design low-complexity measurement matrices and highly reliable signal recovery algorithms.

In recent years, approximate message-passing (AMP) type algorithms have been rapidly developed for signal recovery in compressed sensing \cite{donoho2009message,ma2017orthogonal,rangan2019vector,takeuchi2021bayes,liu2022memory,xie2023massive}. When the entries of $\bf{\Psi}$ are independent and identically distributed (IID), AMP is a low-complexity efficient algorithm for reliable signal recovery. However, when the entries of $\bf{\Psi}$ are correlated, AMP may perform poorly or even diverge. To solve this issue, orthogonal/vector AMP (OAMP/VAMP) is proposed to achieve Bayes-optimal performance for general right-unitarily-invariant matrices $\bf{\Psi}$ in \cite{ma2017orthogonal,rangan2019vector}. However, it is difficult to apply to large-scale systems due to the inherently high-complexity linear minimum mean-square error (LMMSE) estimator. To address this difficulty, a low-complexity convolutional AMP (CAMP) has been proposed \cite{takeuchi2021bayes}, but it struggles to ensure convergence when the condition number of $\bf{\Psi}$ is relatively high. Independently, a memory AMP (MAMP) is developed in \cite{liu2022memory} by replacing the LMMSE of OAMP/VAMP with a memory matched filter, which has been proven to be Bayes-optimal for right-unitarily invariant matrices with low complexity.  

Meanwhile, AMP-type algorithms are widely used for signal detection in multicarrier modulation systems, where ${\bf A}$ denotes the channel matrix and ${\bf \Xi}$ represents the modulation transform matrix. For example, cross-domain/delay-Doppler OAMP (CD/DD-OAMP) detectors are separately proposed for OTFS in \cite{OTFS, DDOAMP}. However, due to the matrix inversion in LMMSE, the CD/DD-OAMP detector struggles to effectively utilize the sparsity of the channel matrix. To address this issue, a DD-MAMP detector is proposed for OTFS, exploiting the sparsity of the equivalent channel matrix in DD domain, but it overlooks the sparser time-domain channels\cite{DDMAMP}. Recently, a cross-domain MAMP (CD-MAMP) detector has been proposed for interleaving frequency division multiplexing (IFDM) in \cite{CDMAMP} to achieve Bayesian optimality with low complexity, which leverages super sparse time-domain channels and right-unitarily invariance in the interleave-frequency-domain channels. Although MAMP achieves Bayesian optimality with low complexity for right-unitarily-invariant measurement matrices and IFDM \cite{liu2022memory,CDMAMP}, high-dimensional measurement matrices in large-scale systems still result in high-complexity challenges in hardware implementation of transceivers. 

To address this challenge, this letter proposes an interleaved block-sparse (IBS) transform, comprising interleaved multiple low-dimensional transform matrices to construct a sparse measurement matrix, thereby reducing the hardware implementation scale and complexity. Meanwhile, an IBS-CD-MAMP estimator is proposed that iteratively performs a memory linear estimator (MLE) in the interleave block-sparse transform (IBST) domain and a non-linear estimator (NLE) in the source domain. Numerical results show that IBS-CD-MAMP operates with lower implementation scale and complexity while preserving performance in IBS-based compressed sensing and IFDM systems.

\section{System Model and Preliminaries}
In this section, we present a linear system model with two application scenarios: compressed sensing and multicarrier modulation. Then, we introduce the CD-MAMP estimator for this system, analyze existing challenges, and highlight the necessity of our proposed methods.
\subsection{System Model}
We consider a large-scale noisy linear system with two constraints:
\begin{equation}
    \label{Eqn:twocon}
    \Gamma: {\bf{y}} = {\bf{A\Xi s} + \bf{n}}, \quad \Phi: s_i\sim P_s, \;\;\forall i,
\end{equation}
where ${\bf{y}} \in {\mathbb{C}}^{M\times 1}$ is an observed vector, ${\bf{A\Xi}}\in {\mathbb{C}}^{M\times N}$ a measurement matrix, $\bf{\Xi}$ a transform matrix, ${\bf{s}}\in {\mathbb{C}}^{N\times 1}$ a vector to be recovered, ${\bf{n}} \backsim \mathcal{C}\mathcal{N}\left ( {0,\sigma }^{2}{\bf{I}}_M\right)$ an AWGN vector, and ${\bf A}$ a rectangular diagonal matrix in compressed sensing or a sparse channel matrix in multicarrier modulation. The number of non-zero elements per row in ${\bf A}$ is $P$ ($P\ll \rm{min}\{M,N\}$). The entries of $\bf{s}$ are IID, i.e., $s_i\sim P_s$. When $\bf{s}$ follows a Gaussian distribution, the optimal solution is the conventional LMMSE estimation. However, for non-Gaussian $\bf{s}$, the problem is generally NP-hard. 

To address this challenge, our goal is to find an MMSE estimate for $\bf{s}$ based on MAMP, which can achieve fast convergence and Bayesian optimality for general right-unitarily-invariant matrices with lower complexity compared to AMP and OAMP/VAMP\cite{liu2022memory}.
% We employ MAMP to address this challenge. Unlike AMP, which is limited to IID measurement matrices, MAMP is capable of handling general right-unitarily invariant matrices. Furthermore, MAMP offers lower complexity compared to OAMP/VAMP, and it demonstrates fast convergence and Bayes optimality \cite{liu2022memory}. Our aim is to find an MMSE estimate of $\bf{s}$. 
The MAMP consists of an MLE and an NLE, involving two local processors $\gamma_t$ and $\phi_t$:
\begin{equation}
    \label{LE_NLE}
    {\rm MLE}:{\bf r}_t=\gamma_t({\bf S}_t), \quad {\rm NLE}:{\bf s}_{t+1}=\phi_t({\bf r}_t),
\end{equation}
where ${\bf S}_t=[{\bf s}_1,...,{\bf s}_t]$ for MAMP, and ${\bf S}_t=[{\bf s}_t]$ for non-memory AMP-type algorithms, represented by OAMP/VAMP. 

\subsection{Application Scenarios}
\subsubsection{Compressed Sensing}
In the context of compressed sensing, where ${\bf\Xi}$ is a partial transform matrix (specifically, ${\bf\Xi}=[\hat{\bf \Pi}_N{\bf{F}}_N]_{1:M}$), and $\bf{s}$ follows a symbol-wise Bernoulli-Gaussian distribution, the system model then unfolds as:
\begin{equation}
    \label{Eqn:CS}
    {\bf{y}} = {\bf{A\Xi s} + \bf{n}}={\bf A}[\hat{\bf{\Pi}}_N{\bf{F}}_N]_{1:M}\bf{s} + \bf{n},
\end{equation}
where ${\bf A}$ is a rectangular diagonal matrix, $[\hat{\bf{\Pi}}_N{\bf{F}}_N]_{1:M}$ consists of ${M}$ rows randomly extracted from the $N$-point normalized FFT matrix ${{\bf{F}}_N \in {\mathbb{C}}^{N\times N}}$, and $\hat{\bf{\Pi}}_N$ is a random permutation matrix for ${\bf{F}}_N$ to ensure the right-unitarily invariance of ${\bf\Xi}$. The compression ratio is denoted by $\delta = M/N$, with $0 < \delta < 1$, i.e., $M<N$.
% implying that $M < N$ in the compressed sensing scenario.
In this context, the signal recovery problem can be effectively addressed using MAMP \cite{liu2022memory}.

\subsubsection{Multicarrier Modulation}
Let ${\bf{x} = \bf{\Xi s}}$ represent the transmit signal in the time domain, where $\bf{s}$ is an information sequence in the source domain. The system can then be expressed as:
\begin{equation}
   \label{Eqn:SM}
    {\bf{y}} = {\bf{A\Xi s} + \bf{n}}= {\bf{Ax} + \bf{n}},
\end{equation}
where $\bf{A}$ denotes the static multipath or mobile time-varying channel matrix, and $\bf{\Xi}$ represents the multicarrier modulation matrix. 
The various values of $\bf{\Xi}$ indicate different multicarrier modulation techniques:
\begin{itemize}
    \item OFDM \cite{OFDM}: In OFDM, the multicarrier modulation matrix is given by the inverse FFT, denoted as ${\bf \Xi}={\bf{F}_N^{\mathrm{H}}}$. Therefore, $\bf{s} = \bf{F}_N\bf{x}$ represents the information sequence in the frequency (F) domain.
    
    \item OTFS \cite{OTFS}: In OTFS, $\bf \Xi = {{{\bf F}}_{J}^{\mathrm{H}}{\otimes}{{\bf I}_K}}$, where $N=KJ$, ${\otimes}$ represents the Kronecker product and ${\bf I}_K$ is the $K$-dimensional identity matrix. In this context, the information sequence $\bf{s}$ resides in the delay-Doppler (DD) domain.
     
    \item AFDM \cite{AFDM}: In AFDM, $\bf \Xi = {{\bf \Lambda}}_{{c}_{1}}^{\mathrm{H}}{\bf F}_N^{\mathrm{H}}{{\bf \Lambda}}_{{c}_{2}}^{\mathrm{H}}$, where ${{\bf \Lambda}}_{{c}_{i}}\triangleq \mathrm{diag}({e}^{-j2\pi c_i n^2},n=0,...,N-1),i=1,2$. Here, $\bf{s}$ is situated in the discrete affine frequency transform (DAFT) domain.
    
    \item IFDM \cite{CDMAMP}: In IFDM, $\bf \Xi = {\hat{\bf{\Pi}}_N}{\bf F}_N^{\mathrm{H}}$, where ${\hat{\bf{\Pi}}_N}$ is a random permutation matrix. Here, $\bf{s}$ is located in the interleave frequency (IF) domain.
\end{itemize}

IFDM has the best BER performance with low modulation complexity \cite{CDMAMP}, which is the focus of this letter. 

\subsection{Cross-Domain Memory AMP Estimator}
\begin{figure}[t]
  \centering 
  \includegraphics[width=7.5cm]{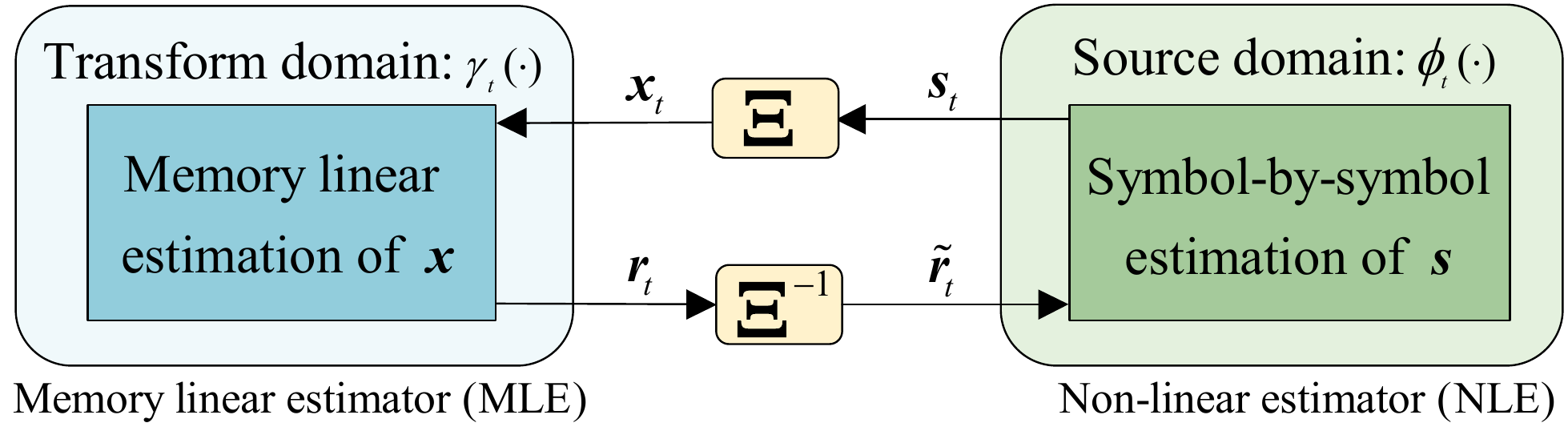}\\ \vspace{-2mm}
  \caption{Graphic illustration of CD-MAMP estimator.}\label{Fig:CD-AMP} \vspace{-5mm}
\end{figure}
The cross-domain MAMP (CD-MAMP) is employed for signal estimation, significantly reducing computational complexity in the iterative process by leveraging the specific structures of ${\bf A}$ and ${\bf \Xi}$. As illustrated in Fig.  \ref{Fig:CD-AMP}, a CD-MAMP estimator involves four processors:
\begin{subequations}\label{LE_NLE2}\begin{align}
    {\rm MLE\  (Transform\ domain)}:&&{\bf r}_t&=\gamma_t({\bf X}_t), 
    \\ {\rm Inverse\ Transform}:&&\tilde{\bf r}_t&={\bf \Xi}^{-1}{\bf r}_t,
    \\{\rm NLE \ (Source\ domain)}:&&{\bf s}_{t+1}&=\phi_t(\tilde{\bf r}_t),
    \\ {\rm Transform}:&&{\bf x}_t&={\bf \Xi}{\bf s}_{t},
\end{align}
\end{subequations}
where $\gamma_t({\bf X}_t)$ is the MLE in the transform domain that mainly depends on $\bf{A}$ and all previous estimations ${\bf X}_t=[{\bf x}_1,...,{\bf x}_t]$. $\phi_t(\tilde{\bf r}_t)$ is an MMSE non-linear symbol-by-symbol estimator in the source domain \cite{liu2022memory} \cite{CDMAMP}. 

The complexity of CD-MAMP per iteration is given by
\begin{align}\label{over O}
    {\cal O}(PN+N+2N\log N),
\end{align}
where ${\cal O}(PN)$ ($P \ll N$) is for the MLE, ${\cal O}(N)$ for the NLE, and ${\cal O}(2N\log N)$ for the fast transform (FT) and inverse FT (i.e., FT and IFT). It is evident that the complexity of CD-MAMP is dominated by ${\cal O}(2N\log N)$,i.e., FT and IFT.
\subsection{Challenge}
In large-scale scenarios, like a linear system with a 4096-bit message sequence, practical systems often struggle to support the ultra-high-dimensional FT computation due to their high complexity and hardware constraints. For example, only a 128-point FT is feasible at a time. This limitation poses significant challenges for the application of the MAMP estimator. Processing the message sequence segment by segment with a 128-bit processor cannot guarantee the right-unitarily invariance, leading to a substantial performance loss. Therefore, it is crucial to develop a low-complexity and low-dimensional transform while maintaining the performance of MAMP. 

\section{Interleaved Block-sparse Transform}\label{Sec:IBST}
This section introduces block-sparse and interleaved block-sparse transforms, followed by a discussion on their fast practical implementation.

\subsection{Block-Sparse Fast Transform (BS-FT)}
We employ block-sparse FT (BS-FT) to address the above challenges. Consider an $N$-bit information sequence in a large-scale linear system, where the practical hardware system with an $N_s$-point FT processor can only process the message segment by segment. The transform matrix ${\bf \Xi}$ is then defined as:
\begin{small}
    \begin{equation}
\!\begin{pmatrix}
 {[\!{\bf{T}}_{1}]}_{1:M_s}&0  &...   &0 \\ 
 0&{[\!{\bf{T}}_{2}]}_{1:M_s}&\ddots  &\vdots  \\ 
 \vdots &\ddots   &\ddots   &0 \\ 
 0      &...  &0    &{[\!{\bf{T}}_{L}]}_{1:M_s} 
\end{pmatrix},
\end{equation}
\end{small}
where $L=N/N_s$ is the number of block-FT matrices, and ${{\bf T}_{l}}$ denotes an ${N_s}$-point FT. We extract the first ${M_s=M/L}$ rows from ${N_s}$ rows in each block. 
Here, each block contains an equal number of rows (${M_s}$). Differences in the number of rows across block-FT matrices decrease overall performance, particularly influenced by the matrix with the fewest rows.

However, segment-by-segment transformation, i.e., BS-FT, prevents the overall signal processing, reduces its randomness, and fails to ensure right-unitarily invariance, resulting in performance degradation. This raises the question: How can we increase the randomness of BS-FT and thereby enhance overall performance?
\begin{figure}
  \centering 
  \includegraphics[width=7cm]{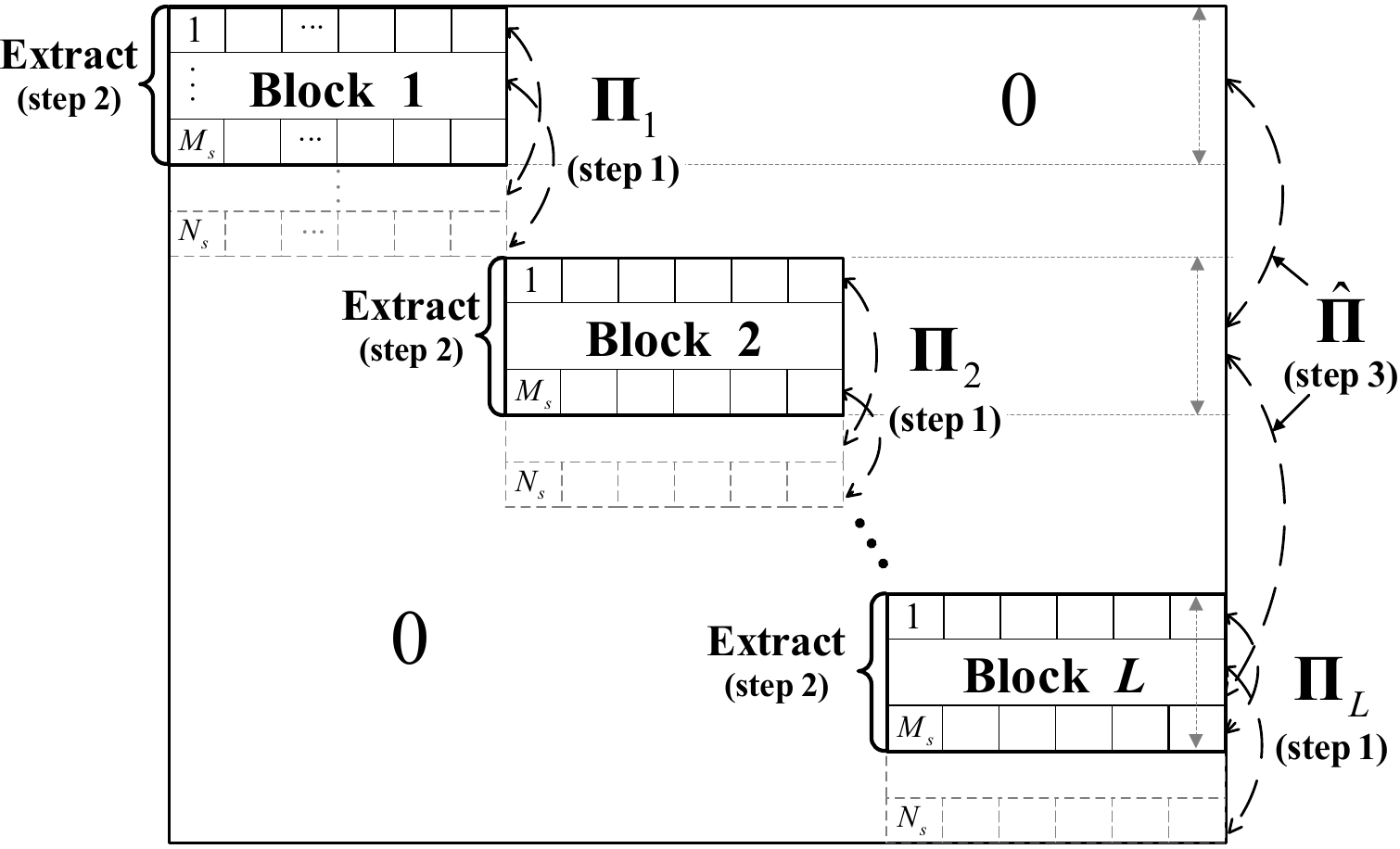}\\ 
  \caption{Graphic illustration for the BW-IBS-FT.}\label{Fig:Final matrix} 
  \vspace{-5mm}
\end{figure}
\subsection{Interleaved Block-Sparse Fast Transform (IBS-FT)}
\begin{figure}[t]
  \centering 
  \includegraphics[width=7cm]{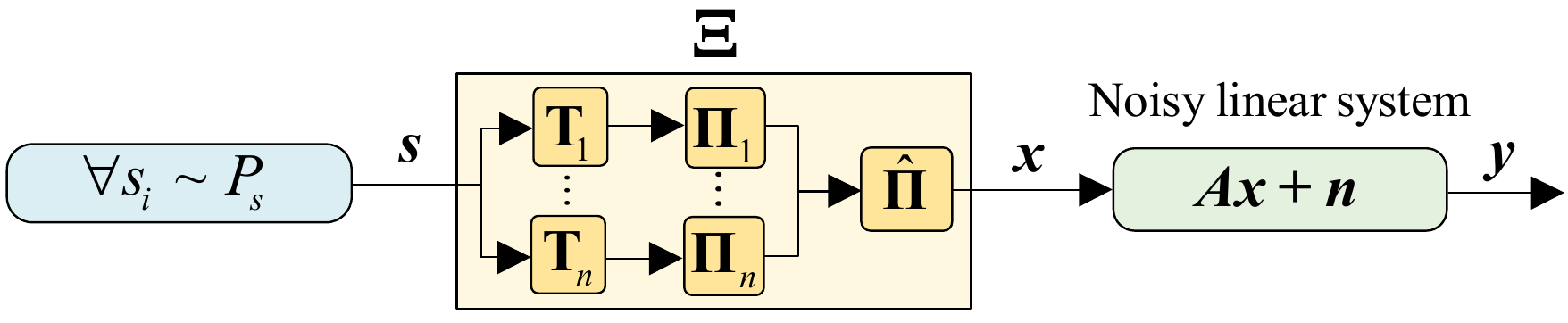}\\ \vspace{-2mm}
  \caption{Graphic illustration for BW-IBS-FT in large-scale noisy linear system.}\label{Fig:linearS} \vspace{-5mm}
\end{figure}

As shown in Fig. \ref{Fig:Final matrix}, to overcome the limitations of BS-FT, we propose a BW-IBS-FT, consisting of block interleaving, block extraction, and whole interleaving as follows:
\begin{itemize}
    \item \emph{Step 1 (Block Interleaving)}: First, the local interleavers  ${{{\bf{\Pi}}_{l}}}$ are employed individually to the $L$ block-FT matrices, each containing an $N_s$-point FT. This step ensures the right-unitarily invariance of each block-FT matrix.
    
    \item \emph{Step 2 (Block Extraction)}: Extract the first ${M_s}$ rows from ${N_s}$ rows in each block, where ${M_s=M/L}$.
    
    \item \emph{Step 3 (Whole Interleaving)}: Finally, a whole interleaver ${\bf{\hat\Pi}}$ is employed to ensure the right-unitarily invariance of the whole transform matrix. 
\end{itemize}
The BW-IBS-FT matrix can be expressed as: 
\begin{small}
    \begin{equation}
{\bf{\hat{\Pi}}}\!\begin{pmatrix}
 {[\bf{\Pi}_{1}\!{\bf{T}}_{1}]}_{1:M_s}&0  &...   &0 \\ 
 0&{{[{\bf{\Pi}}}_{2}\!{\bf{T}}_{2}]}_{1:M_s}&\ddots  &\vdots  \\ 
 \vdots &\ddots   &\ddots   &0 \\ 
 0      &...  &0    &{[{\bf{\Pi}}}_{L}\!{\bf{T}}_{L}]_{1:M_s} 
\end{pmatrix},
\end{equation}
\end{small}
where ${{\bf{\Pi}}_{l}}$ denotes a random interleaving matrix, $[{{\bf{\Pi}}}_{l}{\bf T}_{l}]_{1:M_s}$ the extraction of the first ${M_s}$ rows of ${{\bf{\Pi}}}_{l}{\bf T}_{l}$, and ${\bf{\hat{\Pi}}}$ the whole interleaving matrix. We denote the BW-IBS-FT as: 
\begin{normalsize}
    \begin{equation}\label{Eqn:A1}
     {\bf \Xi}_{\rm BW-IBS} = {\bf{\hat\Pi}}\displaystyle\bigcup_{l=1}^{L}[{\bf{{\Pi}}}_{l}{\bf{T}}_{l}]_{1:M_s}.
\end{equation} 
\end{normalsize}
The large-scale noisy linear system in \eqref{Eqn:twocon} can then be represented in Fig.\eqref{Fig:linearS}. After the BW-IBS-FT, the transmitted signal ${\bf x}$ is:
\begin{equation}  
    {\bf x}=\bf{\Xi} {\bf s}={\bf{\hat\Pi}}(\displaystyle\bigcup_{l=1}^{L}{\bf{{\Pi}}}_{l}{\bf{T}}_{l}){\bf s}.\label{Eqn:IFDM Modu}
\end{equation}

To illustrate the advantages of the proposed BW-IBS-FT, we outline the following schemes for comparison:
\subsubsection{BS-FT (Step 2)} Unlike BW-IBS-FT, BS-FT only involves the extraction in step 2, while there is no local/whole interleaving in steps 1 and 3, i.e., 
\begin{equation}\label{Eqn:BS-FT}
    {\bf \Xi}_{\rm BS}=\displaystyle\bigcup_{l=1}^{L}[{\bf{T}}_{l}]_{1:M_s}.
\end{equation}
\subsubsection{W-IBS-FT (Step 2 + Step 3)}
W-IBS-FT consists of the extraction in step 2 and the whole interleaving for the whole transform matrix in step 3, while there is no local interleaving in step 1, i.e.,  
\begin{equation}\label{Eqn:IBS-FTW}
    {\bf \Xi}_{\rm W-IBS}={\bf{\hat\Pi}}\displaystyle\bigcup_{l=1}^{L}[{\bf{T}}_{l}]_{1:M_s}.
\end{equation}
\subsubsection{B-IBS-FT (Step 1 + Step 2)}
B-IBS-FT consists of local interleaving for the block transform matrices in step 1 and the extraction in step 2, while there is no whole interleaving in step 3, i.e., 
\begin{equation}\label{Eqn: IBS-FTB}
    {\bf \Xi}_{\rm B-IBS}=\displaystyle\bigcup_{l=1}^{L}[{\bf{{\Pi}}}_{l}{\bf{T}}_{l}]_{1:M_s}.
\end{equation}

\begin{table}[t]
\renewcommand{\arraystretch}{1.3}  
\caption{Comparison of Block-Sparse Fast transform Schemes}  
\label{Tab:compare_block}
\centering \footnotesize \setlength{\tabcolsep}{1.5mm}\scalebox{0.88}{
\begin{tabular}{|c|c|c|c|c|}
\hline
Schemes & {Permutation}\ ({$\bf \Pi$}) & Expression & Randomness& MSE Rank \\ \hline
BS-FT & 
None&{\scriptsize $\displaystyle\bigcup_{l=1}^{L}[{\bf{T}}_{l}]_{1:M_s}$} & \begin{tabular}[c]{@{}c@{}}{4}\\ {(Weak)}\end{tabular} & 4 \\ \hline
W-IBS-FT & 
Whole& {\scriptsize${\bf{\hat\Pi}}\displaystyle\bigcup_{l=1}^{L}[{\bf{T}}_{l}]_{1:M_s}$} & 3 & 3 \\ \hline
B-IBS-FT & Block-Wise &{\scriptsize $\displaystyle\bigcup_{l=1}^{L}[{\bf{{\Pi}}}_{l}{\bf{T}}_{l}]_{1:M_s}$} & 2 & 2 \\ \hline 
BW-IBS-FT &\begin{tabular}[c]{@{}c@{}}Block-Wise\\ \& \ Whole\end{tabular} &{\scriptsize ${\bf{\hat\Pi}}\displaystyle\bigcup_{l=1}^{L}[{\bf{{\Pi}}}_{l}{\bf{T}}_{l}]_{1:M_s}$} & \begin{tabular}[c]{@{}c@{}}{1}\\ {(Strong)}\end{tabular} & \begin{tabular}[c]{@{}c@{}}{1}\\ {(Best)}\end{tabular}  \\ \hline
\end{tabular}}\vspace{-5mm}
\end{table} 

Table \ref{Tab:compare_block} provides a comparison of the block-sparse fast transform schemes. The BW-IBS-FT outperforms the other block-sparse fast transform schemes including BS-FT, W-IBS-FT, and B-IBS-FT. As we can see, BW-IBS-FT achieves a favorable trade-off between hardware complexity and performance. For simplicity, we refer to BW-IBS-FT as IBS-FT unless otherwise specified. 
\subsection{Complexity of IBS-FT}
For an $N$-bit message sequence, IBS-FT requires $L$ $N_s$-point FTs, resulting in an overall complexity of $N\log N_s$. Given that the complexity of the original $N$-point FT in \cite{liu2022memory} is $N\log N$, we define the \emph{relative complexity} of the proposed IBS-FT as the ratio of the complexity of IBS-FT to that of the original FT, given by: 
\begin{equation}
    \Theta_{\rm IBS} = {\cal O}(\log_{N}{N_s})={\cal O}(1-\log_{N}{L}),
\end{equation}
where $L=N/N_s$.

\subsection{IBS-FFT and IBS-FWHT}
The proposed IBS-FT can leverage various transforms with fast implementation algorithms, such as the fast Fourier transform (FFT), fast Walsh-Hadamard transform (FWHT), etc.

\subsubsection{IBS-FFT} The IBS-FFT is a specific IBS-FT on the Fourier basis that can be implemented by FFT. In this case, the signal ${\bf{s}}$ (after the FFT) can be depicted as: 
\begin{equation}
    x(n)=\displaystyle\sum_{i=0}^{N-1}s(i){{e}^{-j\frac{2\pi ni}{N}}},
\end{equation}
which involves complex exponential functions, requiring complex arithmetic and substantial hardware resources.

\subsubsection{IBS-FWHT} To address the limitations of FFT, we turn to Walsh-Hadamard Transform (WHT), which utilizes Walsh functions (i.e., a set of orthogonal real functions) that typically take values of $\pm 1$. The signal after WHT is represented as:
\begin{equation}
    x(n)=\displaystyle\sum_{i=0}^{N-1}\{s(i)\displaystyle\prod_{k=0}^{p-1}[R(k+1,i)]^{\left \langle n_k\right \rangle}\},
\end{equation}
where $R(k+1,i)$ is the Rademacher function, and $n_k\in\left\{ 0,1\right\}$ denotes the $k$-th digit of the natural binary code. WHT involves only addition and subtraction with real numbers, avoiding complex number operations. This makes WHT more favorable for hardware implementation and more resource-efficient. 
 
\section{CD-MAMP Estimator for IBS-FT Systems}
\begin{figure}[t]
  \centering 
  \includegraphics[width=8cm]{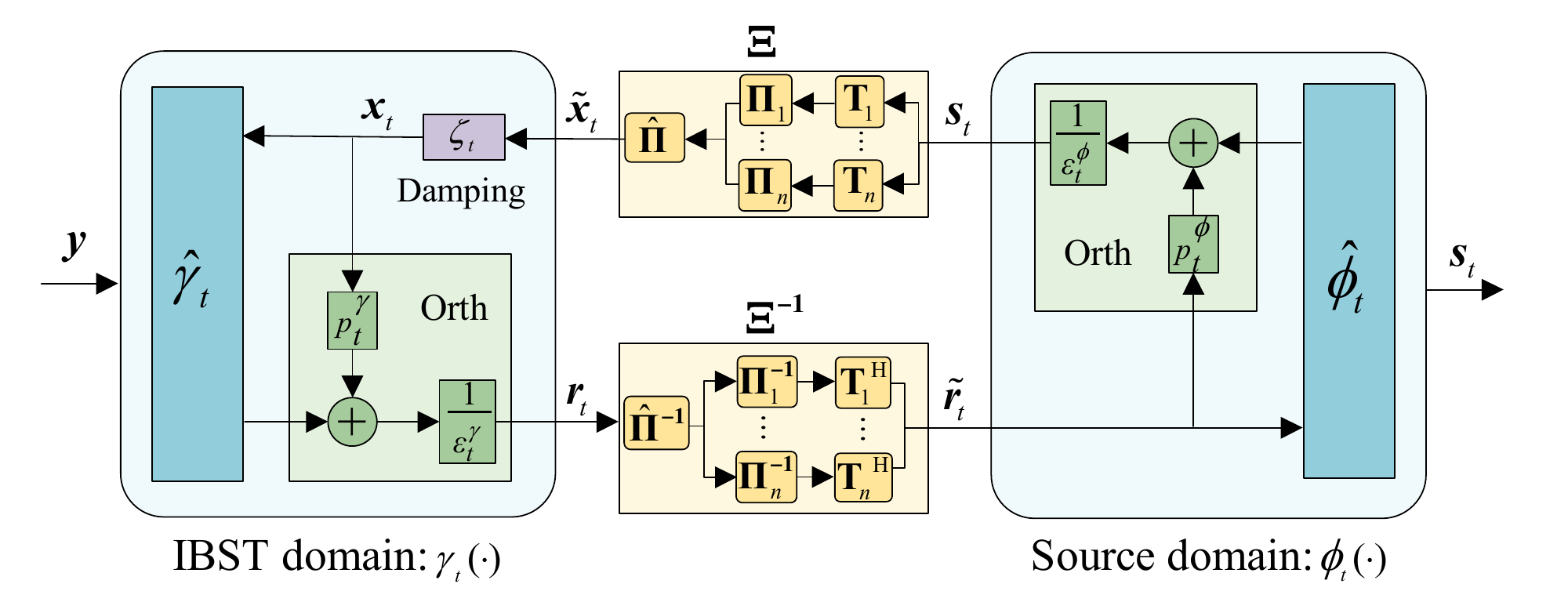}\vspace{-2mm}
  \caption{Graphic illustration for the IBS-CD-MAMP estimator.}\label{Fig:CD-MAMP} 
  \vspace{-5mm}
\end{figure}

As shown in Fig. \ref{Fig:CD-MAMP}, based on the CD-MAMP framework in \eqref{LE_NLE2}, we design an IBS-CD-MAMP estimator for the IBS-FT systems, which consists of an MLE in the IBS transform (IBST) domain, an NLE in the source domain, along with IBS-FT and the inverse IBS-FT (IBS-IFT).

\begin{framed}
\emph{\textbf{IBS-CD-MAMP}}: Define ${\bf X}_t\equiv[{\bf x}_1,...,{\bf x}_t]$. Starting with $t=1$ and ${\bf X}_1=0$,
\begin{subequations}\label{Eqn:CD-MAMP}\begin{align}
{\rm MLE}:&& {\bf r}_t =&\tfrac{1}{{\varepsilon}^\gamma_t}(\bf{A}^{\mathrm H}\hat{\gamma}_t({\bf X}_t)-{\bf p}_t{\bf X}_t), \label{Eqn:CD-MAMP_a}\\
{\rm IBS\mbox{-}IFT}:&& \tilde{{\bf r}}_t=&(\displaystyle\bigcup_{l=1}^{L}{\bf{T}}_{l}{\bf{{\Pi}}}_{l}^{-1}){\bf{\hat\Pi}}^{-1}{\bf r}_t,\label{Eqn:CD-MAMP_b}\\
{\rm NLE}:&& {\bf s}_{t+1} =& \tfrac{1}{{\varepsilon}^\phi_t}(\hat{\phi}_t(\tilde{\bf r}_t)-{ p}_t^\phi\tilde{\bf r}_t),\label{Eqn:CD-MAMP_c}\\
{\rm IBS\mbox{-}FT}:&& \tilde{\bf x}_t=&{{\bf{\hat\Pi}}(\displaystyle\bigcup_{l=1}^{L}{\bf{{\Pi}}}_{l}{\bf{T}}_{l}^\mathrm{H}}){\bf s}_t,\label{Eqn:CD-MAMP_d}\\
{\rm Damping}:  && {\bf x}_t=&[{\bf X}_{t-1},\tilde{\bf x}_t]\cdot{\bf{\zeta}}_t.\label{Eqn:CD-MAMP_e}
\end{align}
\end{subequations}\vspace{-7mm}
\end{framed}\vspace{-2mm}

Since IBS-IFT and IBS-FT have been thoroughly introduced in Section \ref{Sec:IBST}, we will now elaborate on the remaining components: MLE, NLE, and damping.

\emph{1) MLE}: In \eqref{Eqn:CD-MAMP_a}, the normalized parameters $\{{\varepsilon}^\gamma_t\}$ and the orthogonal parameters $\{{\bf p}_t\}$ are utilized to ensure the orthogonality for MAMP. In addition,
\begin{equation}
    \hat{\gamma}_t({\bf X}_t)=\theta_t{\bf B}\hat{\gamma}_{t-1}({\bf X}_{t-1})+ \xi_t({\bf y} - {\bf A}{\bf x}_t),
\end{equation}
where $\{\theta_t, \xi_t\}$ can be optimized to improve the convergence of MAMP, ${\bf B}=\lambda^\dag\bf{I} - {\bf A}{\bf A}^{\mathrm H}$ with $\lambda^\dag=(\lambda_\mathrm{min}+\lambda_\mathrm{max})/2$ with $\lambda_\mathrm{min}$ and $\lambda_\mathrm{max}$ being the minimal and maximal eigenvalues of ${\bf A}{\bf A}^{\mathrm H}$ \cite{liu2022memory}. 

\emph{2) NLE}: In (\ref{Eqn:CD-MAMP}b),  $\{{{\varepsilon}^\phi_t}, { p}_t^\phi\}$ are the normalized and orthogonal parameters in the source domain, which are similar to the $\mathrm{MLE}$, and
\begin{equation}
\hat{\phi}_t(\tilde{\bf r}_t)=\mathrm E\{\bf s|\tilde{\bf r}_t,s_i\sim P_s\}.
\end{equation}

\emph{3) Damping}: ${\bf{\zeta}}_{t}=[{\zeta}_{t,1}, \dots,{\zeta}_{t,t} ]^{\mathrm{T}}$ is a damping vector that can be analytically optimized. For the details, refer to \cite{liu2022memory}. 
 
Compared to the MAMP discussed in \cite{liu2022memory} and the CD-MAMP presented in \cite{CDMAMP}, the proposed IBS-CD-MAMP replaces the conventional FT/IFT with IBS-FT/IBS-IFT. This offers the advantages of a reduced hardware implementation scale and lower complexity, making it particularly well-suited for integration into compressed sensing and IFDM systems aimed at simplifying hardware requirements \cite{CDMAMP}.

\section{Numerical Results}
In this section, we evaluate the performance of our proposed IBS-FT and the IBS-CD-MAMP estimator in compressed sensing and multicarrier modulation systems. 

\begin{figure*}[t]\vspace{-2mm}
  \centering 
  \includegraphics[width=16cm]{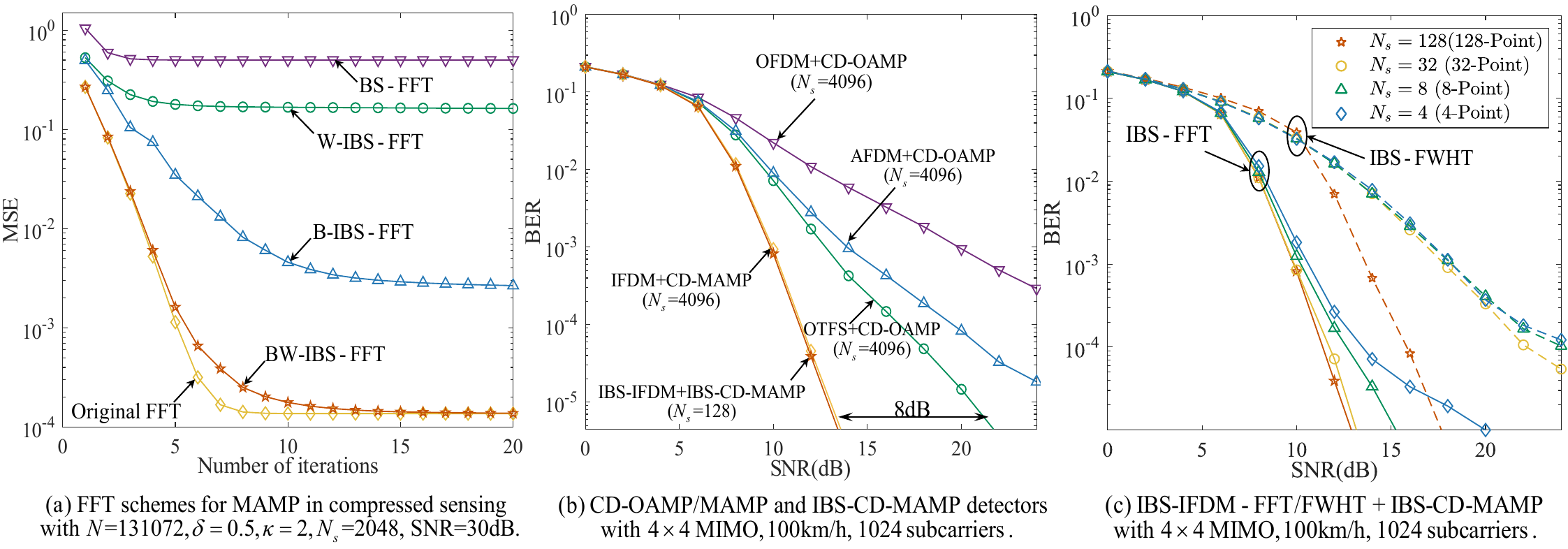}\\ \vspace{-2mm}
  \caption{Comparisons of IBS-FT performance in compressed sensing and multicarrier modulation.}\label{Fig:paper_result} \vspace{-0mm}
\end{figure*}

\emph{1) Compressed Sensing}: We study a large-scale compressed sensing problem with $N=131072$, and $\delta ={M/N}=0.5$. The signal $\bf{s}$ follows a symbol-wise Bernoulli-Gaussian distribution, and the signal-to-noise ratio (SNR) is defined as ${\rm SNR}=1/\sigma^2$. The entries $\{\alpha_i\}$ in ${\bf A}$ are generated by: $\alpha_i/\alpha_{i+1}=\kappa^{1/M}$ for $i=1,\ldots, M-1$ and $\sum_{i=1}^M\alpha_i^2=N$, with $\kappa> 1$ for ill-conditioned $\bf{A}$ (See details in \cite{liu2022memory}). Fig. \ref{Fig:paper_result}(a) shows the MSE versus the number of iterations for FFT schemes in MAMP. Only BW-IBS-FFT with FFT implementation scale $N_s=2048$ converges to the original FFT with $N_s=131072$. Other schemes exhibit worse MSE in MAMP, indicating that BW-IBS-FFT offers superior performance in both MSE and implementation scale.
% \begin{figure}[h]
% \centering 
%   \includegraphics[width=6.5cm]{MAMP_FFT_613.pdf}\\ \vspace{-2mm}
%   \caption{MSE comparisons of FFT schemes for MAMP.}\label{Fig:paper_result1_1} 
% \end{figure}

%Fig. \ref{Fig:paper_result1_2} shows the MSE comparison of BW-IBS-FFT/FWHT with different FT scales $N_s$ for OAMP/MAMP. For simplicity, we refer to BW-IBS-FFT/FWHT as IBS-FFT/FWHT. As $N_s$ decreases, the MSE of OAMP/MAMP in IBS-FFT/FWHT increases, but with lower complexity and smaller hardware implementation size. As shown, IBS-FFT converges to a lower MSE than IBS-FWHT in OAMP/MAMP, but the MSE difference is minimal when $N_s$ is high. Thus, IBS-FWHT has higher computational efficiency and simpler operations but worse MSE performance compared to IBS-FFT. Therefore, when deciding whether to use IBS-FFT or IBS-FWHT in OAMP/MAMP, it is crucial to balance both computational efficiency and MSE performance.
% \begin{figure}[h]
% \centering 
%   \includegraphics[width=6.5cm]{FWHT1_0613.pdf}\\ \vspace{-2mm}
%   \caption{MSE comparisons of BW-IBS-FFT schemes for OAMP/MAMP with $N_s=1024/512$.}\label{Fig:paper_result1_2} 
% \end{figure}

\emph{2) Multicarrier Modulation}: We then compare the performance of advanced estimators for IBS-IFDM and other modulation techniques. We consider a QPSK-modulated MIMO system with mobile time-varying channels, where the number of subcarriers is $N=1024$ ($K=64$, $J=16$ for OTFS). To be fair, we use the same bandwidth for all multicarrier modulations, i.e., the subcarrier spacing is $\triangle f$ kHz for OTFS and OFDM, and $\triangle f/J$ kHz for AFDM and IFDM. Then we consider a carrier frequency of 4 GHz with $\triangle f$=15kHz, a device velocity of 100 km/h, and a channel Doppler shift simulated using the Jakes model. Fig. \ref{Fig:paper_result}(b) compares the BER of CD-OAMP/MAMP and IBS-CD-MAMP. It is clear that the IBS-CD-MAMP for IBS-IFDM ($N_s=128$) achieves the same BER as CD-MAMP for IFDM ($N_s=4096$), both outperforming CD-OAMP for OTFS by 8 dB, AFDM and OFDM by more than 10 dB. Fig. \ref{Fig:paper_result}(c) compares the BER of IBS-FFT and IBS-FWHT in IBS-CD-MAMP. As the implementation scale $N_s$ decreases in IBS-IFDM, both IBS-FFT and IBS-FWHT exhibit degradation in BER performance, with IBS-FFT outperforming IBS-FWHT. 

Table \ref{Tab:compare_complexity} illustrates the relative complexity comparison between IFDM and IBS-IFDM. As the $N_s$ decreases, the relative complexity $\Theta_{\rm IBS}$ significantly reduces. Additionally, the overall relative complexity (w.r.t. IFDM \cite{CDMAMP}), representing the complexity in \eqref{over O}, also declines with lower $N_s$. Experimental results demonstrate that IBS-CD-MAMP effectively reduces complexity and hardware implementation scale compared to the CD-MAMP while preserving BER performance.

\begin{table}[t]
\caption{Comparison of Relative Complexity}
\label{Tab:compare_complexity}\renewcommand\arraystretch{1.2}
\centering \footnotesize \setlength{\tabcolsep}{1.5mm}\scalebox{0.85}{
\begin{tabular}{|c|c|c|c|}
\hline
\begin{tabular}[c]{@{}c@{}}Multicarrier\\ Modulation\end{tabular} & $N_s$ & \begin{tabular}[c]{@{}c@{}}Relative Complexity \\ $\Theta_{\rm IBS}$ \end{tabular}& \begin{tabular}[c]{@{}c@{}}Overall Relative Complexity \\ (w.r.t. IFDM \cite{CDMAMP})\end{tabular} \\ \hline
IFDM & 4096 & 100\% & 100\% \\ \hline
\multirow{4}{*}{IBS-IFDM} & 128 & 58.33\% & 69.69\% \\ \cline{2-4} 
 & 32 & 41.67\% & 57.57\% \\ \cline{2-4} 
 & 8 & 25\% & 45.45\% \\ \cline{2-4} 
 & 4 & 16.67\% & 39.39\% \\ \hline
\end{tabular}}\vspace{-4mm}
\end{table}
\section{Conclusion}
This letter introduces a low-complexity IBS-FT to construct a sparse measurement matrix with a reduced hardware implementation scale. Meanwhile, an IBS-CD-MAMP is proposed to achieve lower implementation scale and complexity while minimizing performance loss for signal estimation in IBS-based compressed sensing and IFDM systems. Therefore, the IBS-FT and IBS-CD-MAMP are extremely promising solutions for the hardware implementation of transmission and signal recovery in large-scale systems.
% for MAMP to deal with the ultra-high-dimensional transform under hardware cost and complexity constraints, which can be efficiently executed in both the Fourier basis and the Walsh-Hadamard basis. Besides, an IBS-CD-MAMP estimator was proposed to reduce the complexity of the CD-MAMP estimator. Experimental results show that the IBS-FT can effectively reduce the complexity of MAMP while maintaining the MSE performance. Besides, the IBS-CD-MAMP estimator for IBS-IFDM has the same BER performance as the CD-MAMP estimator for IFDM with the advantages of lower complexity and smaller hardware implementation scale.

\bibliographystyle{IEEEtran}

\vfill

\end{document}